\documentclass[12pt]{article}
\usepackage{times}
\usepackage{txfonts}
\usepackage{graphics, graphicx}
\usepackage{color}
\usepackage{bbold}
\usepackage{bm}

\renewcommand\sec[1]{\vspace{0.05in}\noindent{{\large\bf{#1}}}
\addtocounter{section}{1}\setcounter{subsection}{0}
}
\newcommand\subsec[1]{\vspace{0.05in} \noindent{\bf{#1}}\addtocounter{subsection}{1}
}

\renewcommand{\b}{\begin{equation}}
\newcommand{\e}{\end{equation}}
\newcommand{\intd}[1]{\int\!\!d#1\,}

\newcommand{\m}{\mbox{\scriptsize m}}

\newcommand{\eq}{\!=\!}
\newcommand{\gt}{\!>\!}
\newcommand{\lt}{\!<\!}

\newcommand{\lte}{\!\le\!}

\begin{document}

%Title Page
 \thispagestyle{empty}
\vspace*{0.5in}
\begin{center}
\begin{Large}
{\bf Dynamics of Random Neural Networks with Bistable Units\\
\vspace*{0.05in}
}
\end{Large}
\vspace*{0.2in}
{\bf Merav Stern$^{1,4}$, Haim Sompolinsky$^{1-3}$ and L. F. Abbott}$^{4,5}$\\
\vspace*{0.1in}
$^1$The Edmond and Lily Safra Center for Brain Sciences\\
$^2$Racah Institute of Physics\\
Hebrew University\\
Jerusalem, Israel\\
\vspace*{0.05in}
$^3$Center for Brain Science\\
Harvard University\\
Cambridge, MA 02138 USA\\
\vspace*{0.05in}
$^4$Department of Neuroscience\\
$^5$Department of Physiology and Cellular Biophysics\\
Columbia University College of Physicians and Surgeons\\
New York NY 10032-2695 USA\\

\vspace*{0.3in}
{\bf Abstract}
\end{center}

We construct and analyze a rate-based neural network model in which self-interacting units represent clusters of neurons with strong local connectivity and random inter-unit connections reflect long-range interactions.  When sufficiently strong, the self-interactions make the individual units bistable. Simulation results, mean-field calculations and stability analysis reveal the different dynamic regimes of this network and identify the locations in parameter space of its phase transitions.  We identify an interesting dynamical regime exhibiting transient but long-lived chaotic activity that combines features of chaotic and multiple fixed-point attractors.

\newpage
\sec{Introduction}

A substantial fraction of the synaptic input to a cortical neuron comes from nearby neurons within local circuits, while the remaining synapses carry signals from more distal locations.  Local connectivity can have a strong effect on network activity (Litwin-Kumar and Doiron, 2012).  In firing-rate models, a cluster of neurons with similar response properties is grouped together, and their collective activity is described by the output of a single unit (Wilson and Cowan, 1972).  Interactions between the neurons within a cluster are represented in these models by self-coupling, that is, feedback connections from a unit to itself, whereas interactions between clusters are represented by connections between units.  Networks consisting of $N$ units with connections chosen randomly and independently have provided a particularly fruitful area of study because they have interesting features and can be analyzed, in the large $N$ limit, using mean-field methods (Sompolinsky et al., 1988).   Self-couplings in the networks that have been studied in this way to date are either non-existent or weak (of order $1/\!\sqrt{N}$).  If these units represent strongly interacting local clusters of neurons, we should include self-coupling of order $1$ in the network model.  For this reason, we consider the properties of firing-rate networks with strong self-interactions.  The remaining interactions, those between units, are taken to be random in our study, reflecting the fact that we are investigating the properties of generic networks, not networks designed to perform specific tasks.

The self-coupling we introduce to represent intra-cluster connectivity, together with the neuronal nonlinearity, can cause the individual units of the network to be bistable.  The random inter-unit connectivity promotes chaotic activity,  as has been previously established (Sompolinsky et al., 1988).  With both forms of connectivity, the networks  we study combine two features normally seen independently, chaotic and multiple-fixed-point attractors.  Our goal is to reveal the different types of activity that arise in networks with self-interacting units and to explore how chaotic and multiple-fixed-point dynamics interact.  We begin by using network simulations to uncover the different dynamic regimes that the network exhibits, and then we use both static and dynamic mean-field methods to determine, in the limit of large network size, the properties of the activity within these regimes and to compute the phase boundaries between them in the space of network parameters.

\sec{The Model and Simulation Results}

The networks we study consist of $N$ units described by activation variables $x_i$, for $i\eq 1, 2, \ldots N$, obeying the equations
\b
\frac{dx_i}{dt} = -x_i + s\tanh(x_i) + g\sum_{j\neq i}^N J_{ij}\tanh(x_j) \, .
\label{netEq}
\e
The second term on the right side of this equation describes the within-cluster coupling, which has a strength determined by the parameter $s$.  The last term on the right side reflects the random cluster-to-cluster interactions.  The elements of the $N\times N$ connection matrix $J$ are drawn independently from a Gaussian distribution with mean $0$ and variance $1/N$, and the parameter $g$ defines the strength of the inter-unit coupling, also known as the network gain.  Note that the form of equation \ref{netEq} implies that time is dimensionless or, equivalently, that it is measured in units of the network time constant.

\begin{figure}[htb]
\centerline{\includegraphics[width= 0.8\textwidth]{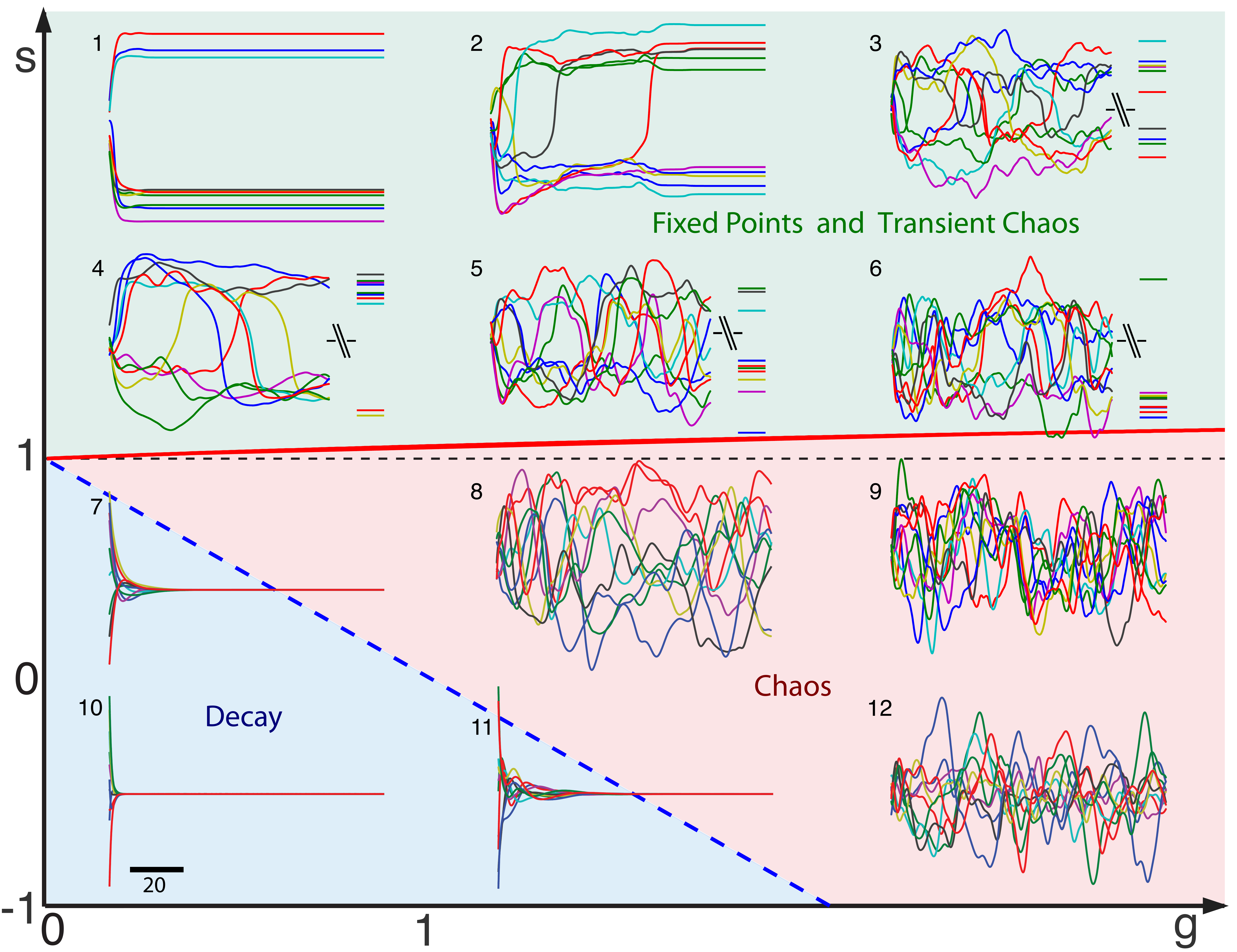}}
{\footnotesize Figure 1.\ Examples of network activity as a function of $s$ and $g$. Each inset shows $x(t)$ for 6 out of 400 network units as a function of time, with its location indicating the values $g$ and $s$ used: for insets 1-12 (in order): $(g,s)\eq$ (0.5,2.5), (1.3,2.5), (2.5,2.5), (0.6,1.5), (1.5,1.5), (2.5,1.5), (0.4,0.4), (1.5,0.5), (2.5,0.5), (0.4,-0.4), (1.2, -0.3), (2.5,-0.5).  The long-dashed line is the boundary between activity that decays to 0 (inserts 7, 10 \& 11) and persistent chaotic activity (inserts 8, 9 \& 12).  The solid curve is the boundary between persistent chaos and what we will show to be transient chaotic activity that ultimately converges to one of many nonzero fixed points  (inserts 1-6).  For inserts 3-6, there is a break in the time axis, reflecting the long time required for convergence to a fixed point. The short-dashed line simply indicates $s\eq1$.}
\label{fig:gsphase}
\end{figure}

The self coupling $s$ and the network gain $g$ determine the network dynamics.  Before considering the full range of values for $s$ and $g$, it is instructive to consider two special cases.  The first is when the self coupling vanishes, $s\eq 0$.  In this case, previous work (Sompolinsky et al., 1988) has shown that, in the limit $N\rightarrow\infty$, the network exhibits chaotic activity when $g\gt 1$ and activity that decays to $0$ when $g\lt 1$.  The second special case is when the network gain vanishes, $g\eq 0$.  In this situation, the units decouple, and each drives its own activity to a fixed point determined by $x \eq s\tanh(x)$.  For $s \lt 1$, the only solution is $x\eq 0$, which is stable, and therefore all unit activity decays to zero from any initial state.  For $s \gt 1$, there are two nonzero stable solutions (the zero solution is unstable) that are negatives of each other.  Thus, in this case the units show bistability and, because they are independent, there are $2^N$ possible stable fixed-point configurations of the network.  Nonzero values of both $s$ and $g$ can give rise to an interesting interplay between chaos and bistability.  

As a preliminary indication of this richness, we investigate the network dynamics over a range of $s$ and $g$ values by computer simulation (figure~1).  In the region below the long-dashed line in figure~1, any initial activity in the network decays to zero.  Above the solid curve, the network exhibits transient irregular activity that eventually settles into one of a number of possible nonzero fixed points.  This settling can take an extremely long time (as we show below, exponentially long in $N$).  In the region between these two curves, the network activity is persistently irregular.  In the following, we will show that both the persistent irregular activity within this region and the transient irregular activity in the region with stable fixed points are chaotic, and we have labelled them as such in figure~1\@.  The region shown with transient chaos and multiple fixed points is a distinctive feature due to the self interaction.

\sec{The Mean-Field Approach}

To determine the type of activity that the network exhibits in different regions of the space of $s$ and $g$ values, we need to characterize solutions of equation \ref{netEq} and evaluate their stability.  For both of these computations, we take advantage of the random nature of the networks we consider.  To analyze stability, we compute the eigenvalues of stability matrices for various solutions using results from the study of eigenvalue spectra of random matrices.  To extract solutions of the network equations, we make use of mean-field methods that have been developed to analyze the properties of network models in the limit $N\rightarrow\infty$, averaged over the randomness of their connectivity (Sompolinsky et al., 1988).  In this section, we provide a brief introduction to the mean-field approach.

The basic idea of the mean-field method is to replace the network interaction term in equation \ref{netEq} (the last term on the right side) by a Gaussian random variable (Sompolinsky et al., 1988) and to compute network properties averaged over realizations of the connectivity matrix $J$.  Because we are averaging over realizations of $J$, all the units in the network are equivalent, so the $N$ network equations \ref{netEq} get replaced by the single stochastic differential equation
\begin{equation}
\frac{dx}{dt} = - x + s \tanh(x) + \eta(t) \, .
\label{eq:dynamicmf}
\end{equation}
We denote the solutions of this equation by $x(t; \eta)$ to indicate that they depend on the particular realization of the random variable $\eta$ used in equation~\ref{eq:dynamicmf}.  If the mean and covariance of the Gaussian distribution that generates $\eta(t)$ are chosen properly and $N$ is sufficiently large, the family of solutions $x(t; \eta)$ across the distribution of $\eta$ will match the distribution of $x_i(t)$ across $i\eq 1, 2, \ldots N$ that solve equation 1, averaged over $J$.  The consistency conditions that assure this require that the first and second moments of $\eta$ match the first and second moments of the interaction term that it replaces.  The first moment of $\eta$ is 0\@.  In the original network model, the average autocorrelation function of the interaction term, averaged over realizations of the random matrix $J$, is
\begin{eqnarray}
C(\tau)&=&\frac{1}{N}\sum_{i=1}^N\left[\left\langle \sum_{j=1}^N J_{ij}\tanh\Big(x_j(t)\Big)\sum_{k=1}^N J_{ik}\tanh\Big(x_k(t+\tau)\Big)\right\rangle\right]\nonumber \\
&=& \frac{1}{N}\sum_{j=1}^N\Bigg\langle \tanh\Big(x_j(t)\Big)\tanh\Big(x_j(t+\tau)\Big)\Bigg\rangle \, ,
\label{netC}
\end{eqnarray}
where the square brackets denote an average of realizations of $J$, the angle brackets denote an average over $t$, and we have used the identity
\begin{equation}
\left[J_{ij} J_{kl}\right] = \frac{1}{N}\delta_{ik}\delta_{jl}    \, .
\end{equation}
The second moment of $\eta$ is $g^2C(\tau)$.  As we will show below, $C(\tau)$ is calculated in the mean-field approach by averaging over $x(t; \eta)$ rather than over the different $x_i$ in the network as in equation~\ref{netC}. This implies that the second moment of $\eta$ is given by
\begin{equation}
\Big\langle \eta(t)\eta(t+\tau)\Big\rangle = \Bigg\langle \tanh\Big(x(t; \eta)\Big) \tanh\Big(x(t+\tau; \eta)\Big) \Bigg\rangle
\label{eq:etaconds}
\end{equation}
where the average is now over realizations of $\eta$.

In the following sections, we use these results to obtain self-consistent mean-field results for both static and dynamic solutions, that depend on the family of solutions $x(t; \eta)$ of equation~\ref{eq:dynamicmf}. When $s\eq 0$, equation \ref{eq:dynamicmf} is linear and can be solved analytically (Sompolinsky et al., 1988).  With nonzero $s$, equation \ref{eq:dynamicmf} must be solved numerically.  We describe procedures for doing this in the following sections, first for the simpler static case, when $x$ and $\eta$ do not depend on time, and then for the more complex dynamic case when they do.

\sec{Analysis of the Fixed Points of the Model}

The phase plot in figure~1 has 3 regions separated by 2 phase boundaries.  As we will see, fixed-point solutions exist in all three of these regions.  Their stability defines the 2 phase boundaries.  In this section, we identify the fixed points, estimate their number, and analyze their stability.

 \subsec{The Zero Fixed-Point}

The trivial solution $x_i\eq 0$ for all $i$ always satisfies equation \ref{netEq}.  To determine whether this solution will robustly appear we must compute its stability.  The stability matrix for equation \ref{netEq} expanded around the zero solution is
\begin{equation}
\label{eq:stabmat}
M_{ij} = (-1 + s) \delta_{ij} + g J_{ij}  \, .
\end{equation}
Because $J$ is a random matrix with variance $1/N$, its eigenvalues, for large $N$, lie in a circle of unit radius in the complex plane (Ginibre, 1965; Girko, 1984; Tao and Vu, 2010).  For $M$, a factor of $g$ scales this radius, and the diagonal terms shift the eigenvalues along the real axis by an amount $-1+s$.  To ensure that all eigenvalues of $M$ have real parts less than zero, so that the zero fixed point is stable, we must therefore require $-1 + s + g \lt 0$.  Thus, the long-dashed line in figure~1 is described by $s\eq 1-g$.

\subsec{Nonzero Fixed Points}

In addition to the zero fixed point just discussed, the model exhibits non-zero fixed points.  We now use the mean-field approach to find solutions corresponding to these non-zero fixed-points and to determine their stability as a function of $g$ and $s$.  Because we are searching for fixed points, the mean field, $\eta$, is a time-independent Gaussian random variable with zero mean and variance $\sigma^2$ to be determined.  The solutions of the static version of equation~\ref{eq:dynamicmf},
\begin{equation}
x - s \tanh(x) = \eta \, ,
\label{eq:staticmf}
\end{equation}
are time-independent functions $x(\eta)$.  For $s\!\neq\! 0$, equation \ref{eq:staticmf} must be solved numerically, and figure~2 reveals that there are three possible solutions when $s\gt 1$ and $|\eta|\lt\eta_{\m}$.  These multiple solutions correspond to multiple fixed points in the original network model.  

\begin{figure}[htb]
\centerline{\includegraphics[width=0.4\textwidth]{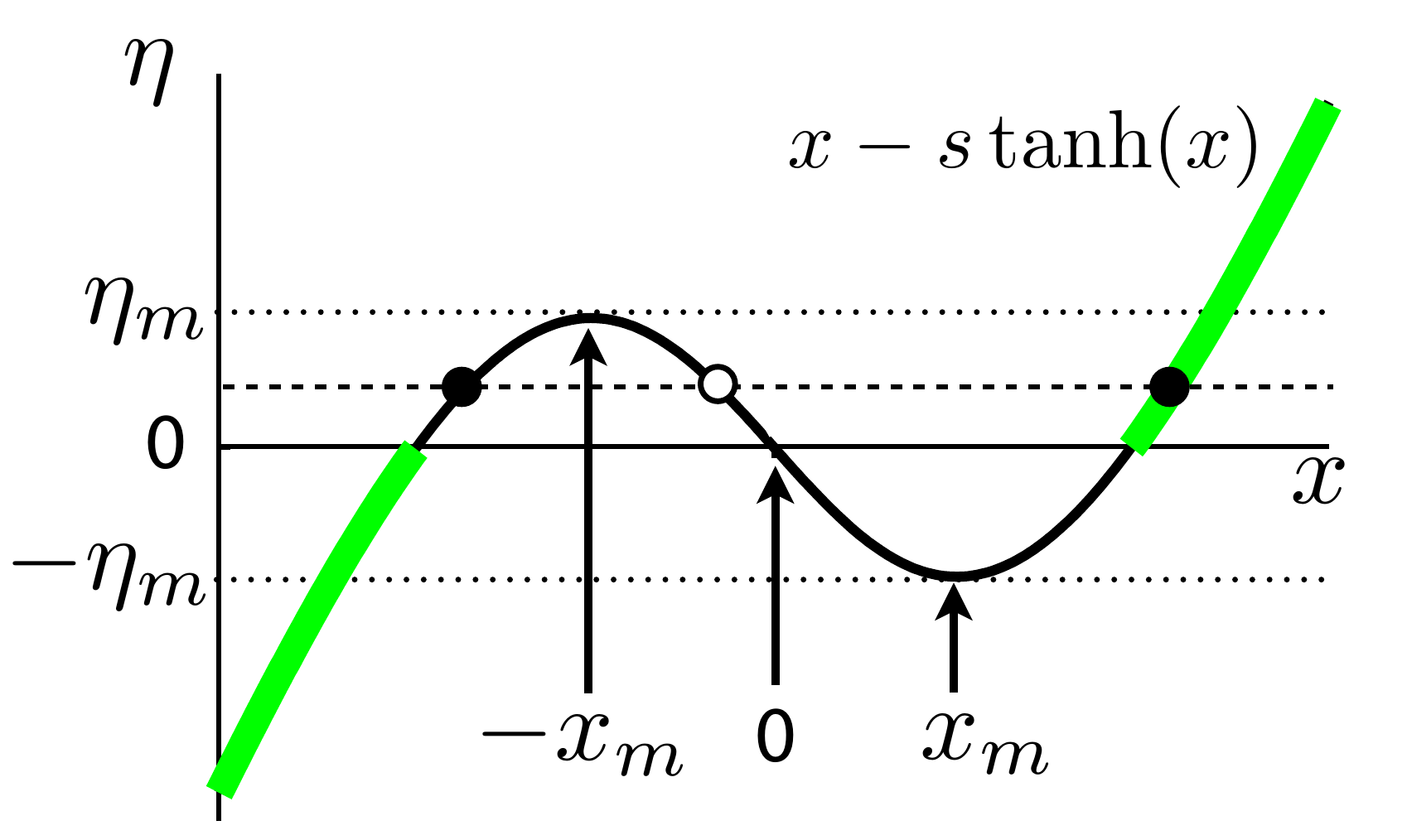}}
{\footnotesize Figure 2.\ Graphical solution of the static mean-field equation \ref{eq:staticmf}.  Solutions $x(\eta)$ are points on the curve $x - s\tanh(x)$ corresponding to a particular value of $\eta$.  The dashed line shows one such $\eta$ value and indicates that there are three possible solutions in the region $-\eta_{\m}\lte \eta \lte \eta_{\m}$ (dots).  Then open circle indicates that solutions along the portion of the curve with negative slope correspond to unstable fixed points (see text).  The arrows show $\pm x_{\m}$, the two local extrema of the function $x - s\tanh(x)$, and $\mp\eta_{\m}$ are its values at these points as indicated by the dotted lines.  Along the transition line between the regions with stable fixed point solutions and persistent chaos (figure 1), the unique stable solutions is restricted to the highlighted portion of the curve where $x(\eta)$ has the same sign as $\eta$.} 
\label{fig:staticMF}
\end{figure}

We are interested in determining the parameter range that supports stable non-zero fixed points and in computing their number.  The stability matrix from equation~\ref{eq:dynamicmf} for a fixed point with values $x_i$, $i\eq1\ldots N$, is
\begin{equation}
M_{ij} = \delta_{ij}\left(-1 + s \Big(1 - \tanh^2(x_j)\Big)\right) + gJ_{ij} \left(1 - \tanh^2(x_j) \right) \, ,
\label{eq:stabM}
\end{equation}
and stability requires that none of its eigenvalues have real parts greater than 0\@.  We can evaluate stability using the mean-field solutions $x(\eta)$, rather than the networks values $x_i$ that appear in equation~\ref{eq:stabM}.  In the limit $N\rightarrow\infty$, the matrix \ref{eq:stabM} has an eigenvalue at the point $z$ in the complex plane if (Ahmadian et al., 2013)
\begin{equation}
\int_{-\infty}^{\infty}\!\! D_{\sigma}\eta\, \left(\frac{g\left(1 - \tanh^2\Big(x(\eta)\Big)\right)}{\left|z + 1 - s \left(1 - \tanh^2 \Big(x(\eta)\Big)\right)\right|}\right)^2 > 1 \, ,
\label{eq:stabz}
\end{equation}
where we use the notation
\b
D_{\sigma}\eta = \frac{d\eta}{\sqrt{2\pi}\sigma}\exp\left(-\frac{\eta^2}{2\sigma^2}\right) \, .
\e
If we ask whether there is an eigenvalue at the point $z\eq 0$, this expression simplifies to
\begin{equation}
Q = \int_{-\infty}^{\infty}\!\! D_{\sigma}\eta\, \left(\frac{g}{\cosh^2\Big(x(\eta)\Big) - s}\right)^2 > 1\, .
\label{eq:atz}
\end{equation}
We use this latter condition below because the inequality~\ref{eq:stabz} does not support isolated eigenvalues (Ahmadian et al., 2013), so stability can be assessed by determining whether or not there is an eigenvalue at $z\eq 0$.  Stability requires $Q\lte 1$, with the edge of stability defined by $Q\eq 1$.

When $s\gt 1$, the expressions in equations~\ref{eq:stabz} and~\ref{eq:atz} are ill-defined as written because $x(\eta)$ is a multivalued function for $|\eta|\lt \eta_{\m}$.  We can eliminate one of the 3 possible solutions of equation~\ref{eq:staticmf} in this range by noting that the denominator of the expression in \ref{eq:stabz} is equal to $z$ plus the slope of the curve drawn in figure~2\@.  Any value of $x(\eta)$ located on a region of this curve with negative slope will cause the denominator to vanish at a positive real value of $z$, indicated the presence of a positive real eigenvalue and instability.  Thus, if we are interested in stable solutions, we can eliminate values of $x(\eta)$ located in the region of negative slope in figure~2, that is, we must require $|x(\eta)|\gt x_{\m}$.  This reduces the number of allowed solutions for $|\eta|\lt\eta_{\m}$ from 3 to 2, one positive, which we call $x_+(\eta)$ and one negative, which we call $x_-(\eta)$.   Because we are interested in evaluating $Q$ of equation~\ref{eq:atz}, we define, in the region where there are two solutions,
\b
f_{\pm}(\eta) = \left(\frac{g}{\cosh^2\Big(x_{\pm}(\eta)\Big) - s}\right)^2  \, .
\e
In the region where there is no ambiguity, we just use $f(\eta)$ to denote this quantity.  Note that, for positive $\eta$, $f_-(\eta)\gt f_+(\eta)$ or, equivalently, $f$ is smaller for the solution with larger $|x(\eta)|$. This means that the solution with larger $|x(\eta)|$, for a given value of $\eta$ in the range $|\eta|\lt \eta_{\m}$, contributes less to $Q$ and hence favors stability.  On the other hand, restricting the solutions to the one with larger $|x(\eta)|$ could eliminate valid stable solutions.

The number of stable fixed points, when they exist, is exponential in $N$, which means that it is exponentially dominated by the configuration of $x(\eta)$ with the different solutions corresponding to stable fixed points.   We define a weighting factor $m(\eta)$ to be the fraction of solutions that are chosen as $x_+(\eta)$ from the two possible values for $x(\eta)$ in this configuration.  Then, $1 - m(\eta)$ is the fraction solutions chosen as $x_-(\eta)$.  With this weighting specified, 
\b
Q = 2\!\!\int_{\eta_{\m}}^{\infty}\!\! D_{\sigma}\eta\, f(\eta)
+ 2\!\!\int_{0}^{\eta_{\m}}\!\! D_{\sigma}\eta\, \Bigg(m(\eta)f_+(\eta) + \Big(1-m(\eta)\Big)f_-(\eta)\Bigg) \, .
\label{QEq}
\e
Note that we have used the $\eta\rightarrow -\eta$ symmetry of the system to express $Q$ in terms of integrals only over the positive range of $\eta$.

The self-consistency condition that determines $\sigma$ (according to the static version of equation~\ref{eq:etaconds}) is also written in terms of the factor $m(\eta)$ as
\begin{equation}
\sigma^2 = 2g^2\!\!\int_{\eta_{\m}}^{\infty}\!\! D_{\sigma}\eta\, \tanh^2\Big(x(\eta)\Big)
+ 2g^2\!\!\int_{0}^{\eta_{\m}}\!\! D_{\sigma}\eta\, \Bigg(m(\eta)\tanh^2\Big(x_+(\eta)\Big) + \Big(1-m(\eta)\Big)\tanh^2\Big(x_-(\eta)\Big)\Bigg) \, .
\label{staticselfconsistent2}
\end{equation}
Finally, the entropy, defined as $1/N$ times the average of the logarithm of the number of fixed points,  is given by counting the number of combinations of $x_+$ and $x_-$ solutions,
\begin{eqnarray}
S_{\mbox{\footnotesize{stable}}} &=& \lim_{N\rightarrow\infty}\frac{2}{N}\!\!\int_{0}^{\eta_{\m}}\!\! D_{\sigma}\eta\, \ln\left(\frac{N!}{\Big(m(\eta)N\Big)!\Big((1-m(\eta))N\Big)!}\right) \nonumber \\
&=& -2\!\!\int_{0}^{\eta_{\m}}\!\! D_{\sigma}\eta\, \Bigg(m(\eta)\ln\Big(m(\eta)\Big) + \Big(1 - m(\eta)\Big)\ln\Big(1 - m(\eta)\Big)\Bigg) \, .
\label{entropy}
\end{eqnarray}

To complete the mean-field calculation, we need to determine the weighting function $m(\eta)$.  We do this by imposing stability on the solutions being integrated in equation~\ref{entropy}.  The entropy is exponentially dominated by solutions at the edge of stability, so we constrain $Q$ to the value 1, rather than imposing the inequality $Q\lte 1$\@.  We then chose $m(\eta)$ to be the function that provides the maximum contribution to the entropy subject to the constraint $Q\eq 1$.  Introducing the Lagrange multiplier $\lambda$ to impose this constraint, we maximize $S + \lambda Q$ (using equations~\ref{entropy} and~\ref{QEq}) with respect to $m(\eta)$, obtaining
\b
m(\eta) = \frac{\exp\Bigg(\lambda \Big(f_+(\eta) - f_-(\eta)\Big)\Bigg)}{1 + \exp\Bigg(\lambda\Big(f_+(\eta) - f_-(\eta)\Big)\Bigg)} \, .
\label{mEq}
\e
Substituting this expression into equation~\ref{entropy}, we see that computing the entropy require the determination of two variables, $\sigma^2$ and $\lambda$.  These are computed numerically by simultaneously solving equation~\ref{staticselfconsistent2} and the condition $Q\eq 1$, using equations~\ref{QEq} and~\ref{mEq}.  

\begin{figure}[htb]
\centerline{\includegraphics[width=\textwidth]{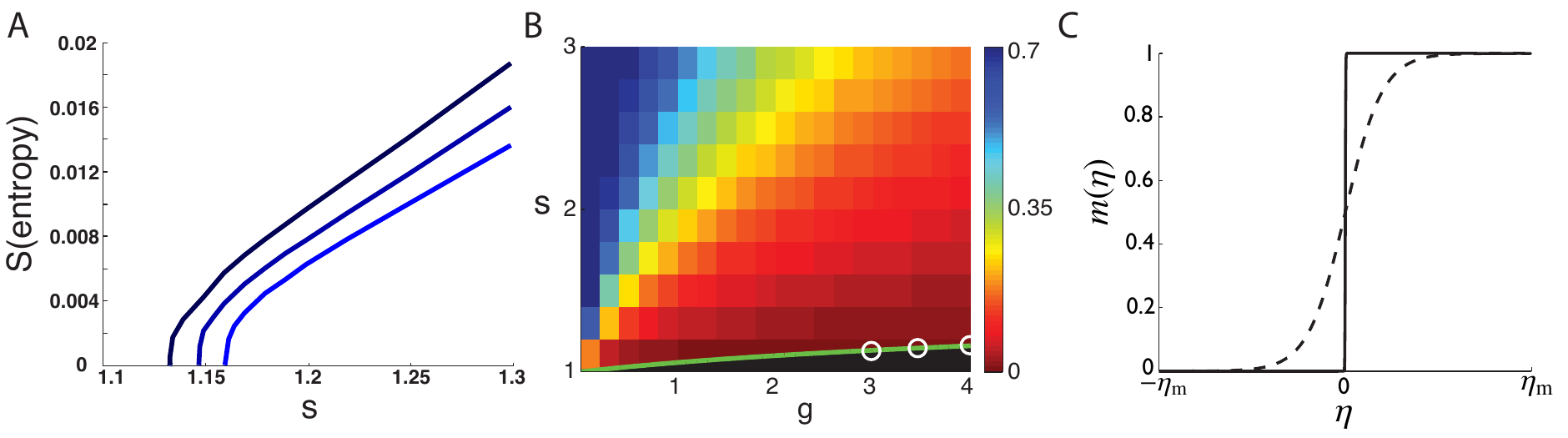}}
\footnotesize{Figure 3.\ The entropy as a function of $s$ and $g$, which is $1/N$ times the average (over $J$) of the logarithm of the number of stable fixed points in the network model.  In both plots, the entropy goes to 0 at the values of $s$ and $g$ corresponding to the phase transition between the nonzero fixed-point and persistent chaotic regions in figure 1\@.  A) The entropy as a function of $s$ for, from top to bottom curves, $g\eq 3, 3.5$ and 4.\@  B) The entropy over a range of $s$ and $g$ values represented by colors.  The green line shows where the entropy reaches 0\@. The white circles indicate the results obtained from the zero intercepts of the curves in A\@. Far from the transition line $m(\eta) \rightarrow 0.5$ for all $|\eta| \lt \eta_{\m}$ and the entropy can be estimated as $\ln(2) \int_{-\eta_m}^{\eta_m} D_{\sigma}\eta$. C) The weighting function $m(\eta)$ near the transition (solid curve; $s \eq 1.133$ and $g\eq 3$) and away from the transition (dashed curve; $s \eq 1.135$ and $g\eq 3$).}
\label{fig:numOfFixedPoints}
\end{figure}

The transition curve between the nonzero-fixed-point and chaotic regions in figure~1 is the set of $s$ and $g$ values for which $Q\eq 1$ and the entropy of stable solutions vanishes (i.e.\ the number of stable solutions approaches 0).  Figure 3A reveals three sets of values for which this occurs.  Recall, that choosing the solution with larger $|x(\eta)|$ decreases $Q$, enhancing stability.  At the transition, $\lambda\rightarrow -\infty$, and $m(\eta)$ is therefore a step function (figure 3C). This restricts all the solutions for $\eta\gt 0$ to be positive and all the solutions for $\eta\lt 0$ to be negative.  The set of $s$ and $g$ values for which $S\rightarrow 0$ is the solid curve in figure 3B, which is given approximately by $s_c(g) \approx 1 + 0.157 \ln{(0.443g + 1)}$.  This is the transition line between the chaotic and transient-fixed-point regions in figure 1\@.  Results for the entropy away from the transition line are indicated by colors in figure 3B.

\sec{Analysis of the Dynamics of the Model}

We now examine the dynamics of the network model .  We begin by studying solutions of the dynamic mean-field equations and using them to compute the average autocorrelation function of the network units.  We then examine other properties of the network dynamics.

\subsec{Autocorrelation}

To study network dynamics, we return to the time-dependent mean-field equation, equation~\ref{eq:dynamicmf}.  The mean-field $\eta(t)$ is a random variable with zero mean and correlation function 
\b
\Big\langle \eta(t)\eta(t+\tau)\Big\rangle = g^2C(\tau)\, ,
\label{etaEq}
\e
where the angle brackets denote averages over the distribution that generates $\eta(t)$, and $C(\tau)$ is to be determined self-consistently (Sompolinsky et al., 1988).  It is easiest to express $\eta(t)$ in terms of its Fourier transform $\tilde\eta(\omega)$ and the Fourier transform of $C(\tau)$, $\tilde C(\omega)$, as
\begin{equation}
\tilde\eta(\omega) = g \tilde C^{1/2}(\omega) \xi(\omega) \, .
\label{etaTEqn}
\end{equation}
Here, $\xi$ is a complex random variable with real and imaginary parts chosen independently, and independently for each discrete value of $\omega$, from a Gaussian distribution with zero mean and variance 1/2\@.  This assures that equation~\ref{etaEq} is satisfied.  The self-consistency condition that determines $C(\tau)$, which equates the correlation of the mean field to the average auto-correlation of the network interaction term in equation~\ref{netEq}, is then expressed in terms of a functional integral over $\xi(\omega)$ as
\begin{equation}
C(\tau) = \int \!\!\frac{\delta \xi(\omega)}{\sqrt{2\pi}}\int \!\!\frac{\delta \xi^*(\omega)}{\sqrt{2\pi}}\, \exp\left(-\intd{\omega}|\xi(\omega)|^2\right) 
\tanh\Big(x(t; \eta)\Big)\tanh\Big(x(t+\tau; \eta)\Big) \, ,
\label{selfconsistent2}
\end{equation}
with $\eta$ given by the Fourier transform of equation~\ref{etaTEqn} and $x(t, \eta)$ by equation~\ref{eq:dynamicmf}.  Equation~\ref{selfconsistent2} is a self-consistency condition because its right side depends on $C(\tau)$ through equation~\ref{etaTEqn}.

We use an iterative approach to solve the dynamic mean-field equations.  We start by making an initial guess for the function $C(\tau)$, perform a discrete Fourier transform $C(\tau)\rightarrow \tilde C(\omega)$, and use this in equation~\ref{etaTEqn}.  We then compute $\eta(t)$ by inverse discrete Fourier transformation and solve equation \ref{eq:dynamicmf} to obtain $x(t)$.  Computing $x(t)$ for many different draws of $\xi(\omega)$, we compile a large set of solutions that allows us to compute $C(\tau)$ as a Monte-Carlo approximation of the integrals in equation~\ref{selfconsistent2}. Starting with this new $C(\tau)$, instead of our initial guess, we repeat the entire procedure, obtain yet another $C(\tau)$, and iterate until the average across iterations of $C(\tau)$ converges. To check against previous calculation, we have verified that we obtain the same results as in Sompolinsky et al.\ (1988) for $s\eq 0$.

\begin{figure}[htb]
\centerline{\includegraphics[width=0.8\textwidth]{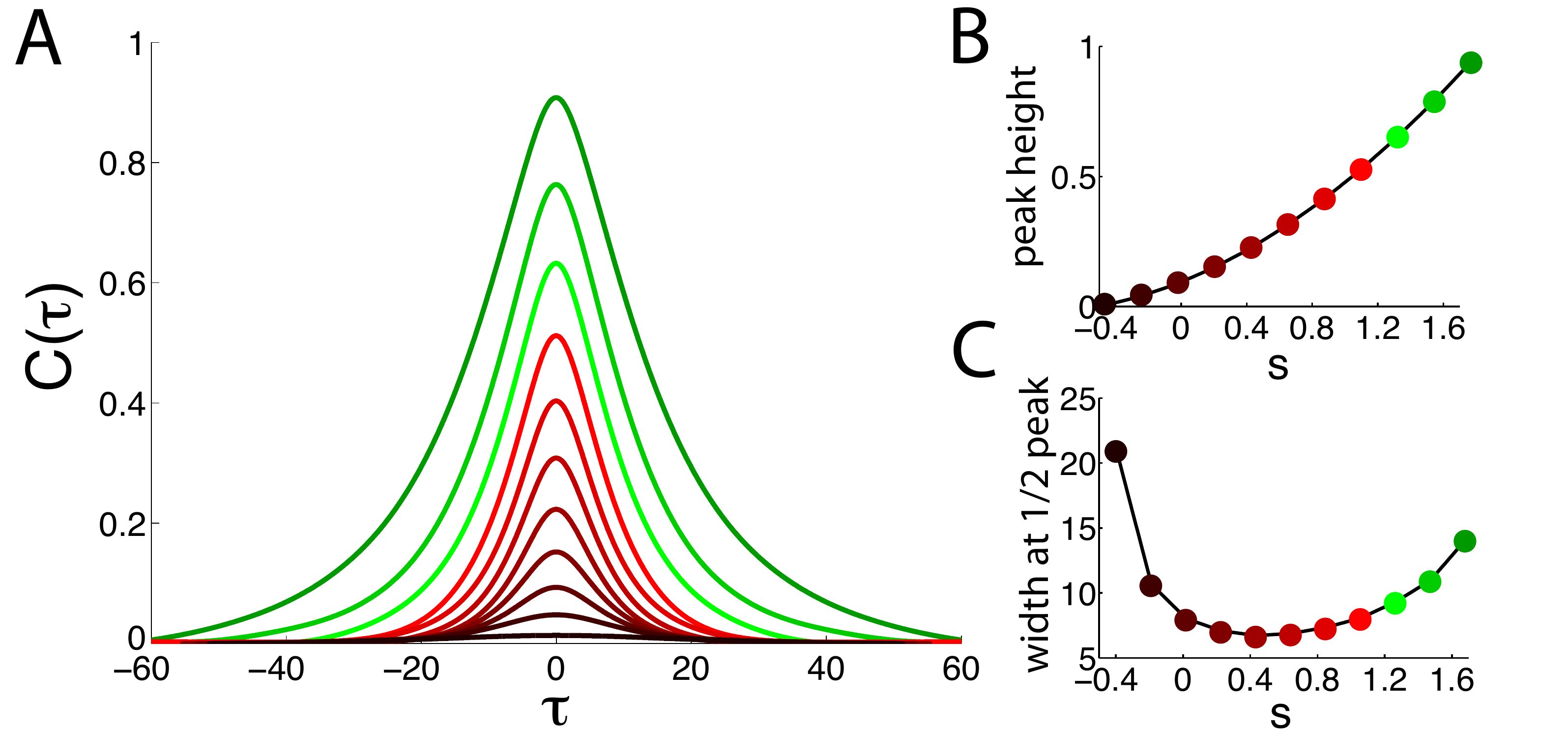}}
{\footnotesize Figure 4.\ Autocorrelation functions for different $s$ values and $g\eq 1.5$.  A) $C(\tau)$ for, from the top to the bottom curve, $s$ ranging from 1.6 to -0.4 in steps of 0.2\@. B) Peak heights of the curves in A\@.  C) Widths at 1/2 peak for the curves in A\@.}
\label{fig:autocShapes}
\end{figure} 

There are no self-consistent solutions of the dynamic mean-field equations in the region where the zero fixed point is stable, but such solutions exists everywhere above the transition line where the zero fixed point becomes unstable  (figure 1).  The shape of the autocorrelation function (figure 4A) varies continuously across the phase diagram, with no discontinuity at the transition between the regions that do and do not support stable nonzero fixed points (figure~1).  The peak height, $C(0)$, increases steadily as a function of either $g$ or $s$ (figure 4B) until it saturates at 1\@.  This reflects the increased cross- or self-coupling driving the units to saturation.  The width of the autocorrelation shows a more interesting non-monotonic dependence (figure 4C).  As expected, the width diverges at the phase transition between the chaotic and zero-fixed-point regions (left side of the plot in figure 4C).  It also diverges for large $s$ and small $g$ (right side of the plot in figure 4C).  

In the region with stable nonzero fixed points (top of figure~1), we have thus obtained two mean-field solution, one static and one dynamic, suggesting the coexistence of stable non-zero fixed points and irregular time-dependent activity in the limit $N\rightarrow\infty$.  To understand how this limiting behavior arises, we study numerically the relationship between these two types of solutions for finite $N$.   

\subsec{Lifetime of the Transient Activity}

\begin{figure}[htb]
\centerline{\includegraphics[width=0.5\textwidth]{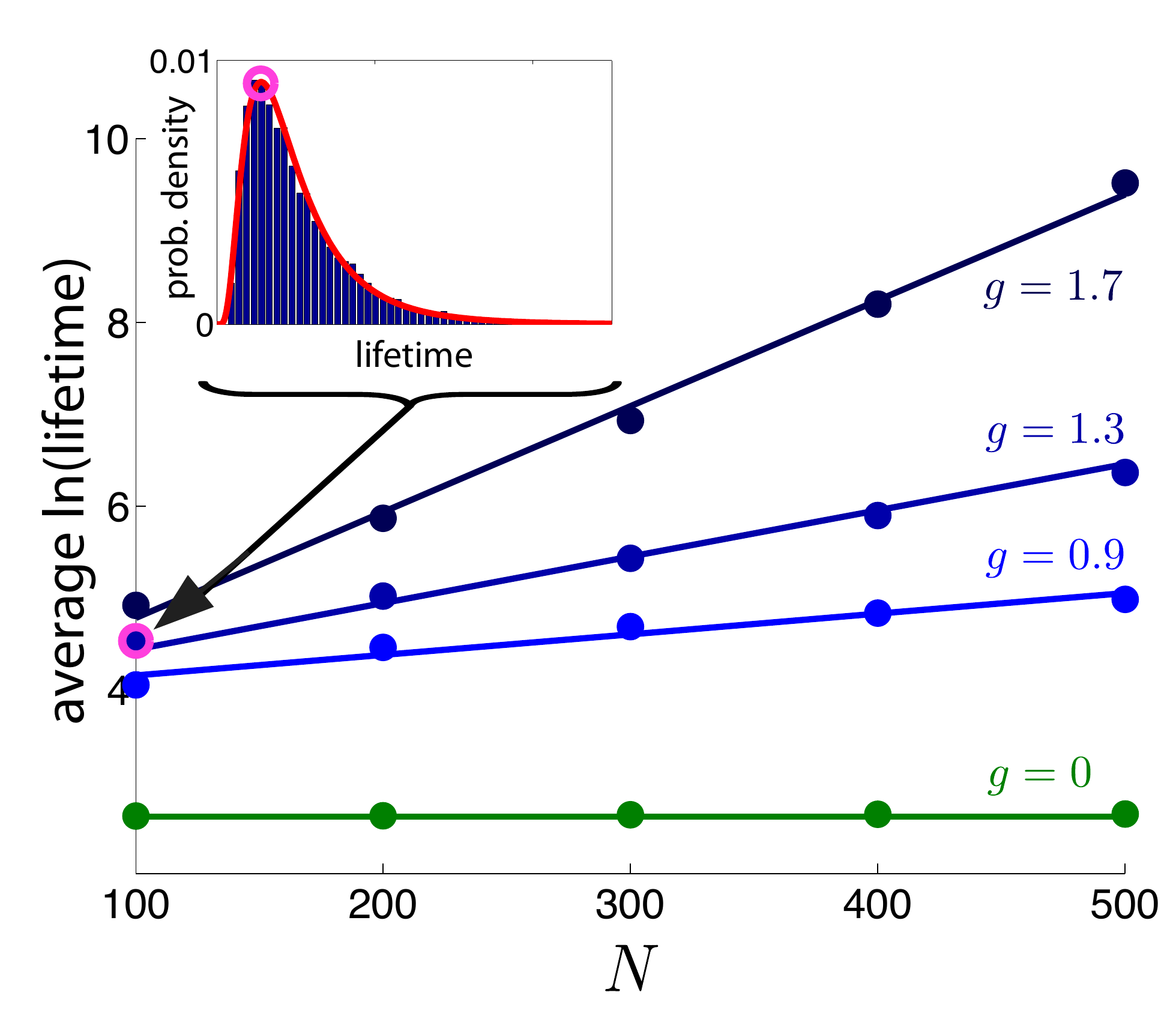}}
{\footnotesize Figure 5.\ Exponential dependence on the network size $N$ of the lifetime of the transient activity in the region with stable nonzero fixed points.  Inset: Distribution of times to reach a fixed point for $N\eq 100$, $s\eq 2.3$ and $g\eq 1.3$, shown with bars.  The curve is a fit to a log-normal distribution.  Main figure: The average of the logarithm of the time to reach a fixed point plotted as a function of $N$, for different $g$ values.  In all these examples, $s\eq 2.3$.  For $g\eq 0$ the units are decoupled and the lifetime is independent of $N$.  For $g\gt 0$, the lifetime is exponential in $N$.}
\label{fig:convByN}
\end{figure}

As shown in figure~1, activity arising from typical initial conditions in the region with stable nonzero fixed points exhibits irregular fluctuations that ultimately decay to one of the stable fixed points.  Results for the lifetime of this transient dynamic activity for different values of $g$ and networks of different sizes are presented in figure~5\@.  The lifetime depends on the initial state of the network, which was chosen randomly, and, for small networks, on the realization of $J$.  We ran 10,000 trials with different draws of $J$ and different initial conditions to obtain a distribution of lifetimes for the transient activity in the region with non-zero fixed points.  This distribution is log-normal (inset, figure~5).  We then computed the average of the logarithm of the lifetime for different $s$ and $g$ values (using 100 trials in each case).  As can be seen in figure~5, the average log-lifetime is linear in the size of the network, and it increases more rapidly with $N$ as $g$ is increased.  The average log-lifetime divided by $N$ and the entropy follow roughly inverse patterns (not shown).  This makes sense because the smaller the number of stable fixed points the longer it should take for the network to find one of them.  In conclusion, we find that the coexistence of static and dynamic states in the mean-field analysis corresponds to the $N\rightarrow\infty$ limit of a transient fluctuating state that transitions to a non-zero fixed point after a time that grows exponentially with $N$.

\subsec{Maximum Lyapunov Exponents}

\begin{figure}[htb]
\centerline{\includegraphics[width=3in]{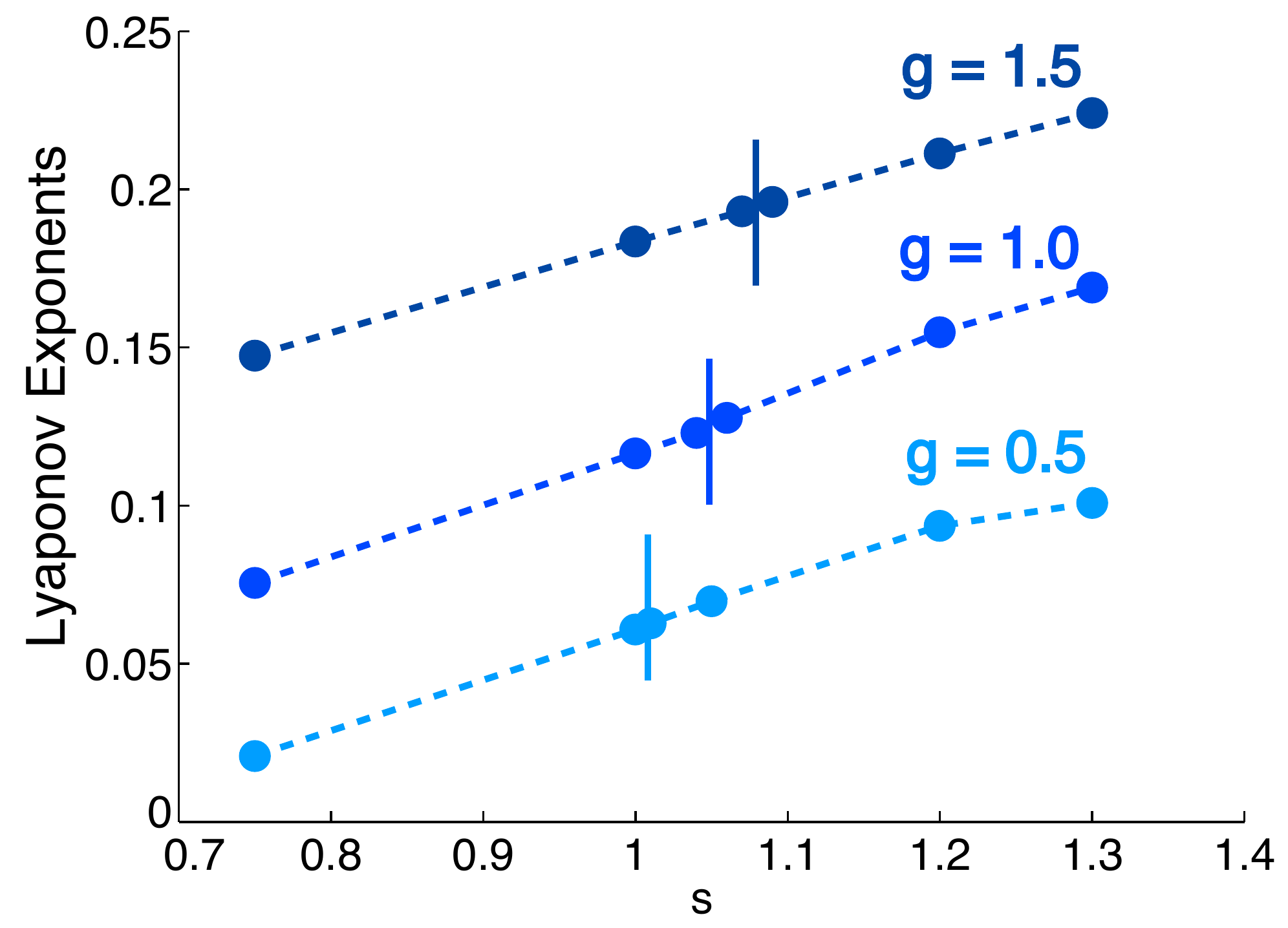}}
{\footnotesize Figure 6.\ Maximum Lyapunov exponents as a function of $s$ for three different $g$ values.  The vertical lines indicate where the transition from persistent chaos (to the left of these lines) to stable fixed points (to the right) occurs. The maximal Lyanpunov exponents vary continuously and smoothly through this transition.}
\label{fig:Lev2}
\end{figure}

In this subsection, we show numerical results for the largest Lyaponov exponents over a range of $g$ and $s$ values in regions with both persistent and transient irregular activity (figure~1).  The long lifetime of the transient activity in the region with nonzero fixed points for finite $N$ allows us to analyze its properties numerically.  In particular, we made sure to use large enough networks so the calculation of the Lyaponov exponent converged before the network reached a fixed point. The largest Lyaponov exponent is positive in both of these regions, indicating the exponential sensitivity to initial condition typical of chaos. The largest Lyapunov exponent increases smoothly as a function of both $s$ and $g$ with no indication of any discontinuity at the transition between the persistent and transient regions (figure~6).  This suggests that there is no sharp distinction between these two forms of chaotic activity, other than their long-term stability.  Rather, as supported by our mean-field results on the correlation function, characteristics of the chaos change continuously across the phase diagram. 

\subsec{Bimodality}

To further characterize the nature of the chaotic activity, we simulated networks exhibiting both transient and persistent chaos and extracted distribution of $x$ values over time and network units.  As seen in figure~7, these show bimodality  that starts within the persistent chaotic region and become more apparent in the region where stable fixed points exist.  Although bistability of individual isolated units requires $s\gt 1$, bimodality appears for values of $s$ well below 1\@. 

\begin{figure}[htb]
\centerline{\includegraphics[width=0.9\textwidth]{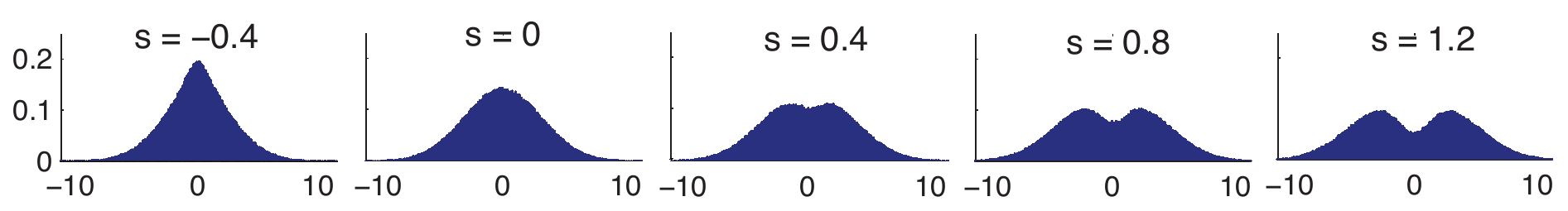}}
\footnotesize{Figure 7.\ Normalized histograms of unit $x$ values for different $s$, with $g\eq 3.5$ and $N\eq 1000$. Run time was $20,000$, and $x$ values were sampled at intervals of $50$ to avoid temporal correlations.  The width of the distributions increases with $s$, and bimodality first becomes apparent between  
$s\eq 0.2$ and $s \eq 0.3$ (not shown).  The transition to the region of transient chaos occurs at $s\eq 1.15$ for this value of $g$.}
\label{fig:bimodality}
\end{figure}

Bimodality, as seen in the histograms of figure~7, arises from the shape of the curve in figure~2. The resulting quasi-bistable behavior can be seen by plotting $\tanh(x)$ as a function of time (figure~8A).  Especially for $s\gt 1$ and small $g$, the activity is characterized by relatively infrequent flips between fluctuations about the two $x$ values where $-x + s \tanh(x)\eq 0$ ($x\eq \pm x_0$), corresponding to $\tanh(x)$ near 1 or -1\@,  with a  log-normal distribution of inter-flip times (figure~8B). The average time between flips shrinks as a function of $g$ and grows as a function of $s$ (figure~8C).  For $s \eq 0$, the inter-flip times, or equivalently times between zero-crossings, follow an exponential distribution. Between small and large value of $s$, the inter-flip distribution changes continuously from exponential to log-normal.

\begin{figure}[htb]
\centerline{\includegraphics[width=0.6\textwidth]{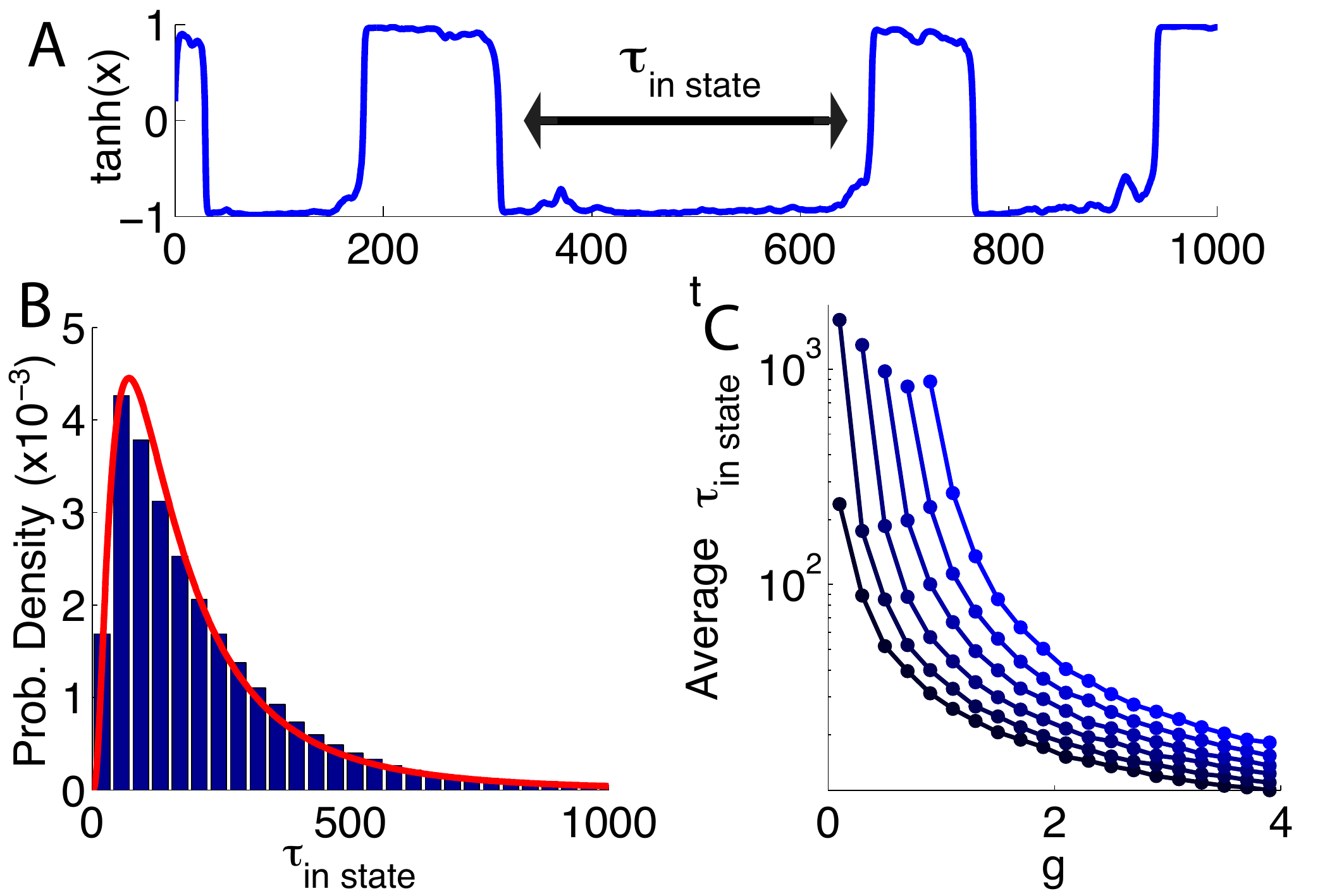}}
{\footnotesize Figure 8.\ A) Typical activity of one unit in the region where stable fixed points exist ($s\eq 1.6$, $g\eq 0.7$). $\tau_{\mbox{\scriptsize{in state}}}$ is defined as the time between flips of the unit between states that fluctuate near $\tanh(x)\eq \pm 1$.  B) Distribution of $\tau_{\mbox{\scriptsize{in state}}}$ values extracted as in A from all the units.  C) The average $\tau_{\mbox{\scriptsize{in state}}}$ for different values of $g$ and $s$, note the logarithmic scale of the vertical axis. From top to bottom, the curves correspond to $s$ ranging from 2 to 1 in steps of 0.2\@.}
\label{fig:flips}
\end{figure}

The flipping of units between quasi-stable states due to network fluctuations may appear similar to the well-studied problem in which a bistable system is perturbed by noise.  A unique feature of this system, however, is that the correlation time of the ``noise'', which is actually the result of chaotic fluctuations, is on the same order as the time between flips.  Thus, the self-consistency condition relates the increase in the width of the correlation function at large $s$ (figure 4C) to the increase in the average time between flips seen in figure~8C\@.

\sec{Discussion}

The network model we have studied interpolates between chaotic networks (for $s$ near zero) and networks with large numbers of stable fixed points (for large $s$), with an intermediate region in which the activity shares features of both.  The intermediate activity ranges from patterns dominated by approximate fixed points that are destabilized by chaotic fluctuations to chaotic activity with bimodal activity distributions.  In the former case, the chaotic activity acts as a form of colored noise, the ``color" induced by its correlations, and drives sign changes in the baseline around which the chaotic fluctuations occur.  This form of activity, dominated by flip-like transitions, ultimately terminates when the network finds a true dynamic fixed point, but this occurs over time periods given by a log-normal distribution with a mean that depends exponentially on the size of the network.

We can envision two types of applications of clustered networks.  First, the long-time-scale dynamics of the flip-like activity might be harnessed through learning algorithms for tasks requiring processing or coherence over long times.  Second, the system could be used as a quasi-stochastic ``noise" source with a tunable spectrum, which could drive internal network states producing a realization of a Hidden-Markov model.  For example, the flips shown in figure~8A have the characteristics of log-normal distributed random events, although they are, of course, actually deterministic.  We leave such applications to future work.

\vspace*{0.4in}
\subsec{Acknowledgements}
We are grateful to Omri Barak for helpful discussions and to Yashar Ahmadian for providing results on random matrix spectra. Research was supported by the Gatsby Charitable Foundation (M.S., H.S. and L.F.A.), the James S. McDonnell Foundation (H.S.), and the Swartz Foundation and NIH grant MH093338 (L.F.A.).

\newpage
\sec{References}
\begin{list}{}{
\setlength{\leftmargin}{0.2in}
\setlength{\parsep}{\parskip}
\addtolength{\parsep}{-8pt}
\setlength{\listparindent}{-0.2in}}
\item \mbox{ }\vspace{-0.35in}

Ahmadian, Y., Fumarola, F. \& Miller, K.D. (2013) Properties of networks with partially structured and partially random connectivity.\  arXiv:1311.4672.

Ginibre, J. (1965) Statistical ensembles of complex, quaternion, and real matrices.\ J.\ Mathematical Phys.\ 6: 440-449.

Girko, V.L. (1984) The circular law.\ Theory Probab.\ Appl.\ 29: 669-679.

Litwin-Kumar, A. \& Doiron, B. (2012) Slow dynamics and high variability in balanced cortical networks with clustered connections.\ Nat.\ Neurosci.\ 15:1498-1505.

Sompolinsky, H., Crisanti, A. \& Sommers, H.J. (1988) Chaos in Random Neural Networks.\ Phys.\ Rev.\ Lett.\ 61:259-262.

Tao, T. \& Vu, V. (2010) Random matrices: Universality of ESD and the Circular Law.\  Annals of Probability 38: 2023-2065.

Wilson, H.R. \& Cowan, J.D. (1972) Excitatory and inhibitory interactions in localized populations of model neurons.\ Biophys.\ J. 12:1-24.

\end{list}
\end{document}